%% file: paper.tex
\documentclass[conference]{IEEEtran}
\IEEEoverridecommandlockouts

\input{configs.tex}

\input{macros.tex}

\begin{document}

\date{}

\title{SPIDER: Fuzzing for Stateful Performance Issues in the ONOS Software-Defined Network Controller}

\author{
{\rm Ao Li}\\
aoli@cmu.edu\\
Carnegie Mellon University\\
Pittsburgh, USA
\and
{\rm Rohan Padhye}\\
rohanpadhye@cmu.edu\\
Carnegie Mellon University\\
Pittsburgh, USA
\and
{\rm Vyas Sekar}\\
vyass@cmu.edu\\
Carnegie Mellon University\\
Pittsburgh, USA
} 


\maketitle

\input{files/abstract.tex}


\input{files/introduction.tex}

\input{files/background.tex}
\input{files/overview.tex}

\input{files/detailed-design.tex}
\input{files/implementation.tex}
\input{files/experiments.tex}

\input{files/related-work}
\input{files/discussion}
\bibliographystyle{IEEEtran}

\bibliography{references}

\end{document}

%% file: configs.tex
\usepackage{tikz}
\usepackage[figure,vlined,linesnumbered]{algorithm2e}

\usepackage{enumitem}
\usepackage{mathpartir}
\usepackage[frozencache]{minted}
\usepackage{mathbbol}
\usepackage{pifont}
\usepackage{amsmath}
\usepackage{tikz}
\usepackage[htt]{hyphenat}
\usepackage{filecontents}
\usepackage{pgfplotstable}
\usepackage{multirow}
\usepackage{url}
\usepackage{hyperref}
\usepackage{amssymb}
\usepackage{nopageno}
\usepackage{microtype}
\IEEEtriggeratref{47}

\usepackage{booktabs} 
\hypersetup{                    
  colorlinks,
  linkcolor={green!80!black},
  citecolor={red!70!black},
  urlcolor={blue!70!black}
    }

\usepackage[font=bf]{caption}
\newcommand{\tightcaption}[1]{\caption{#1}}

\usepackage{xspace}

\newcommand{\xmark}{\ding{55}}

\SetKwInOut{Input}{Input}
\SetKwInOut{Output}{Output}
\SetKw{Continue}{continue}
\setminted{
baselinestretch=0.7,
fontsize=\scriptsize,
linenos,
baselinestretch=1,
breaklines,
xleftmargin=17pt,
escapeinside=??
}

%% file: macros.tex
\setlist[itemize]{topsep=0pt,leftmargin=*,itemsep=1pt}
\setlist[enumerate]{topsep=0pt,leftmargin=*,itemsep=1pt}

\newcounter{packednmbr}

\newcommand{\sysfull}{{\sc{Full}}}
\newcommand{\sysempty}{{\sc{Single}}}

\newcommand{\systemname}{{\sc{Spider}}\xspace}

\newcommand{\fdependency}{\ensuremath{\mathit{Dep}}}

\newcommand{\fevent}{e}
\newcommand{\feventlist}{\sigma}

\newcommand{\fservice}{S}

\newcommand{\stateful}{stateful\xspace}
\newcommand{\performance}{performance\xspace}

\newcommand{\services}{services\xspace}

\newcommand{\statedependency}{service\xspace dependency\xspace}

\newcommand{\blackbox}{black-box\xspace}

\newcommand{\Q}{Q}

\newcommand{\mypara}[1]{{\smallskip \noindent {\bf #1:}}\xspace}

\newcommand{\myparagraph}[1]{{\smallskip \noindent {\bf #1}}\xspace}

\newcommand{\SPIsShort}{SPIs\xspace}
\newcommand{\SPIShort}{SPI\xspace}
\newcommand{\Section}{\S}

%% file: files/abstract.tex
\begin{abstract}
Performance issues in software-defined network (SDN) controllers can have serious impacts on the performance and availability of networks. In this paper, we consider a special class of SDN vulnerabilities called stateful performance issues (SPIs), where a sequence of initial input messages drives the controller into a state such that its performance degrades pathologically when processing subsequent messages. Uncovering SPIs in large complex software such as the widely used ONOS SDN controller is challenging because of the large state space of input sequences and the complex software architecture of inter-dependent network services. We present SPIDER, a practical fuzzing framework for identifying SPIs in this setting. The key contribution in our work is to leverage the event-driven modular software architecture of the SDN controller to (a) separately target each network service for SPIs and (b) use static analysis to identify all services whose event handlers can affect the state of the target service directly or indirectly. SPIDER implements this novel dependency-aware modular performance fuzzing approach for 157 network services in ONOS and successfully identifies 10 new performance issues. We present an evaluation of SPIDER against prior work, a sensitivity analysis of design decisions, and case studies of two uncovered SPIs.
\end{abstract}


%% file: files/introduction.tex
\section{Introduction}


Software-defined networking is increasingly adopted in wide-area, data center, and enterprise networks~\cite{sdnsurvey}. In contrast to traditional networks where routers and switches run both the routing (i.e., control plane) and forwarding (i.e., data plane), SDN logically decouples the control and data plane tasks. To this end, SDN introduces a  {\em controller} (e.g., ONOS~\cite{onos})
that communicates with network devices (routers and switches) through a configuration protocol (e.g., OpenFlow~\cite{openflow}).

The SDN controller performs its tasks based on an internal state that it maintains; this state is updated based on messages received from network hosts and switches and eventually used to configure the entire network. Given the critical role that the SDN controller plays, vulnerabilities in the controller can lead to undesirable outcomes impacting the overall performance, security, and availability of the network~\cite{sdnsecuritysurvey,sdnsecurity2,sdnsecurity1}. 

However, finding vulnerabilities in the SDN controller is not trivial. For example, Open Network Operating System (ONOS) is a leading open-source SDN controller used by many large network providers such as Comcast and AT\&T~\cite{onos}. ONOS contains 150+ network services that communicate with each other asynchronously. Researchers have developed several specialized analyses to identify vulnerabilities such as memory-safety issues, protocol race conditions, and configuration issues in SDN controllers such as ONOS~\cite{sdnracer, d2c2, conguard, delta}.



 In this paper, we consider a new class of vulnerabilities in ONOS, which we call
 \emph{\stateful \performance issues} (\SPIsShort). First, \SPIsShort are performance issues that lead to excessive resource consumption when processing inputs (i.e., OpenFlow messages). Such issues in an SDN controller can severely compromise the network's availability. \SPIsShort can only be triggered after the SDN controller has reached a specific internal state while processing other messages. 
 
Identifying \SPIsShort is challenging because it involves finding a sequence of messages that first puts the SDN controller in a vulnerable state and then triggers a costly operation. At a high level, this is a challenging search-space exploration problem due to a combination of algorithmic and system factors. First, we need to consider a large input search space of long {\em sequences} of OpenFlow messages of interest. The second issue is the large code base and non-trivial software architecture: ONOS has tens of thousands of lines of code comprising hundreds of network services with complex dependencies between them. The third challenge stems from the semantics of \SPIsShort: we need to capture the dependencies between inputs and internal states and identify which state-input combinations induce high resource consumption. 

 We present {\systemname}, a system for identifying \SPIsShort in the ONOS SDN controller. At its core, {\systemname} uses performance fuzzing~\cite{perffuzz, slowfuzz} to automatically generate inputs that maximize execution cost. {\systemname} addresses the aforementioned challenges by implementing a novel \emph{dependency-aware} \emph{modular performance fuzzing} framework. 
 
 Our key observation is that ONOS uses an event-based modular software architecture, where network services communicate with each other using asynchronous \emph{events}. Events are first triggered by incoming OpenFlow messages. Network services subscribe to one or more event types; their event handlers can update their internal state and/or fire other events.
 
 \systemname{} generates \emph{sequences} of internal events to trigger an \SPIShort; that is, where the last event in the sequence exacerbates performance in some service $S$. Our \emph{key insight} in making this scalable is that the only events relevant to such an \SPIShort are those whose processing may directly or indirectly affect the internal state of $S$. \systemname{} leverages this insight in the following way. First, we focus on analyzing one service at a time with the goal of triggering an \SPIShort{} in just that service. Second, when targeting a service $S$, we use \emph{static analysis} to identify inter-service dependencies. Finally, our performance fuzzer uses the dependency information to generate event sequences that \emph{only} contain events that may affect the state of $S$. Such a \emph{dependency-aware modular analysis} allows {\systemname} to reduce the search space without sacrificing fidelity.

For event generation, we borrow the idea from Zest~\cite{zest}, utilizing type-specific generator functions to represent and mutate well-formed inputs. For most event types, these generators can be synthesized automatically from type definitions. However, for approximately 10\% of event types critical to many services, we use handcrafted generators to enhance fidelity.

We use {\systemname} to analyze all 157 services in the ONOS SDN controller. {\systemname} flags 11 potential \SPIsShort, of which 10 are true positives and 9 depend on complex state interactions. We classify these issues based on the capabilities/scenarios required for triggering them and on their impact. The most serious identified vulnerabilities include (a) a malicious host can degrade the SDN controller's performance by cumulatively increasing the cost of processing an OpenFlow message in an unbounded way, and (b) a vulnerability in the topology service leads to worst-case exponential performance, which can be triggered by a compromised network switch. Our experiment also shows that the \SPIShort enables an attacker to reduce the throughput of the controller down to 1Mb/s after sending 4000 spoofed ARP packets at low frequency (10 pkts/s) while only controlling one vulnerable host in the network.

We evaluate our design decisions by comparing {\systemname} with three baseline implementations, including a monolithic SDN fuzzer (Delta~\cite{delta}), a variant of \systemname without dependency information (\sysfull), and a variant of \systemname that analyzes services in isolation (\sysempty). We run separate fuzzing campaigns for all three variants of \systemname{} for each of the 157 services, fixing the budget in terms of fuzzing time and with repetitions ($\sim$1.6 CPU years). Compared to \systemname{}'s 10 true positives, \sysfull{} identifies only 1, and \sysempty{} identifies 2; Delta triggers 3 issues, though isolating the inputs is non-trivial. Our results indicate that {\systemname} is uniquely effective in identifying stateful performance issues in ONOS.

To summarize, this paper makes the following contributions:

\begin{itemize}
    \item We identify a new class of SDN vulnerabilities called \emph{stateful performance issues} (\SPIsShort).
    \item We propose \systemname{}, a novel \emph{dependency-aware modular performance fuzzing} technique for identifying \SPIsShort{} in an event-based software architecture.
    \item We use {\systemname} to implement fuzzers for 157 services in the ONOS SDN controller.
    \item We identify 10 unique performance issues in ONOS and provide detailed case studies for two of the \SPIsShort.
    \item We present a thorough evaluation and compare {\systemname} with three baseline implementations.
\end{itemize}

%% file: files/background.tex
\section{Background and Problem Definition}
\label{sec:background}

\begin{figure}[!tb]
\begin{minted}{Java}
public class ARPService {
  private Map<IpAddress,MacAddress> addressMap;
  private Map<IpAddress,MacAddress> getAddressMap(){
    // Generate a shallow copy of addressMap 
    // by iterating over each entry in the map.
    Map<IpAddress,MacAddress> copy = new HashMap<>();
    for (Map.Entry entry: addressMap.entrySet()) { ?\label{fig:castor:copy}?
      copy.put(entry.getKey(), entry.getValue()); ?\label{fig:castor:write}?
    }
    return copy;
  }
  public void add(IpAddress ip, MacAddress mac) {
    addressMap.put(ip, mac);
  }
  public MacAddress lookup(IpAddress ip) {
    return getAddressMap().get(ip);
  }
  public void packetHandler(OFPacketIn packetIn) {
    Ethernet payload = packetIn.getPayload(); ?\label{fig:castor:payload}?
    if (payload instanceof ARP) {
      ARP arp = (ARP) payload;
      if (arp.opCode == 0x1 || arp.opCode == 0x2) { ?\label{fig:castor:check}?
        if (lookup(arp.ip) == null) { ?\label{fig:castor:lookup}?
          this.add(arp.ip, arp.mac); ?\label{fig:castor:add}?
} } } } }
\end{minted}
\vspace{-0.2cm}
 \tightcaption{Simplified view of \texttt{ARPService} in ONOS, illustrating a stateful performance issue. The \texttt{lookup} function
  triggered by \texttt{OFPacketIn}, performs an $\mathcal{O}(n)$ operation w.r.t. the size of \texttt{addressMap}.
  }
  \label{fig:castor}
\vspace{-0.4cm}
\end{figure}





In software-defined networks (SDNs), the SDN controller, a central node, governs routers and switches using control messages. These devices handle data packets within the network. ONOS~\cite{onos}, a popular open-source Java-based SDN controller, processes input messages from routers and switches and produces corresponding output messages or actions.




ONOS consists of a list of {\em\services}. Each service performs specific network functions, such as processing LLDP packets, and can be dynamically loaded and unloaded. Services register event handlers, maintain local state, and alter states upon event processing. When a service state changes, it may generate and dispatch events delivered to other subscriber services. For example, the LLDP service processes LLDP packets and dispatches topology events if a device is connected or disconnected. Similarly, the Flow service implements logic related to flow rules, listens to the topology events, and updates its internal state.






In this paper, we focus on \emph{\stateful \performance issues} (\SPIsShort) in these services.  Such issues can be a serious concern for critical infrastructures since they can introduce Denial of Service~\cite{Kandoi15, Zhang16, Wang15} or induce subtle tail latency~\cite{deantail}.  
Triggering an \SPIShort involves two phases:
First,  a sequence of inputs drives the system
to a vulnerable state. Then, a specific input consumes an
excessive amount of compute resources.

 \SPIsShort are different from two classical types of potential vulnerabilities explored in the literature. First,   in contrast to  \emph{stateless}  performance issues, where a single input leads to an amplified response (e.g.,~\cite{perfsyn,freezingtheweb}), stateful issues entail a complex sequence of events.  Second, in contrast to stateful security issues related to \emph{protocol state}~\cite{snooze, pulsar, polyglot},  \SPIsShort target the state of internal data structures in the ONOS.
 Although \SPIsShort have been studied in other settings (e.g., databases~\cite{torpedo}), to the best of our knowledge, this has not been explored in the context of SDN controllers.

\SPIsShort are difficult to catch with traditional pre-deployment software testing or in runtime system profiling. First,  the issue may not be revealed in the profiling data from normal runs, as the system may not reach a vulnerable state. Second, a misconfiguration (or attack) can slowly build the state over time (e.g., by infrequently adding ARP records in the example above), and remain undetected until the final trigger.

\myparagraph{Illustrative example.}  Figure~\ref{fig:castor} presents a real issue we 
discovered in the  ONOS \texttt{ARPService}. The service processes
\texttt{OFPacketIn} events with ARP payloads
and stores the mapping between IP and MAC addresses.
\texttt{packetHandler} is an event handler
which processes all \texttt{OFPacketIn} events corresponding to
OpenFlow packets. The  \texttt{OFPacketIn} event may cause the service
to first look up ARP records (Line~\ref{fig:castor:lookup}) and
add a record to the \texttt{addressMap} if the record is missing (Line~\ref{fig:castor:add}).
Unfortunately,  the \texttt{lookup} function has a subtle performance issue. \texttt{lookup}
calls \texttt{getAddressMap()} to get a shallow copy of the \texttt{addressMap} instead of
querying \texttt{addressMap} directly (Line~\ref{fig:castor:copy}). This
 leads to $\mathcal{O}(n)$ operation with respect to the
size of \texttt{addressMap} each time \texttt{lookup} is called.
Note that \texttt{OFPacketIn} events can be triggered by data-plane ARP
packets; e.g., a misconfigured or malicious host can send spoofed ARP packets to increase the  \texttt{addressMap} size $n$, and
each message will subsequently trigger an $\mathcal{O}(n)$ computation in terms of its size. 

%% file: files/overview.tex
\section{Solution Overview}
\label{sec:overview}

\begin{figure}[!tb]
    \centering
    \includegraphics[width=0.47\textwidth]{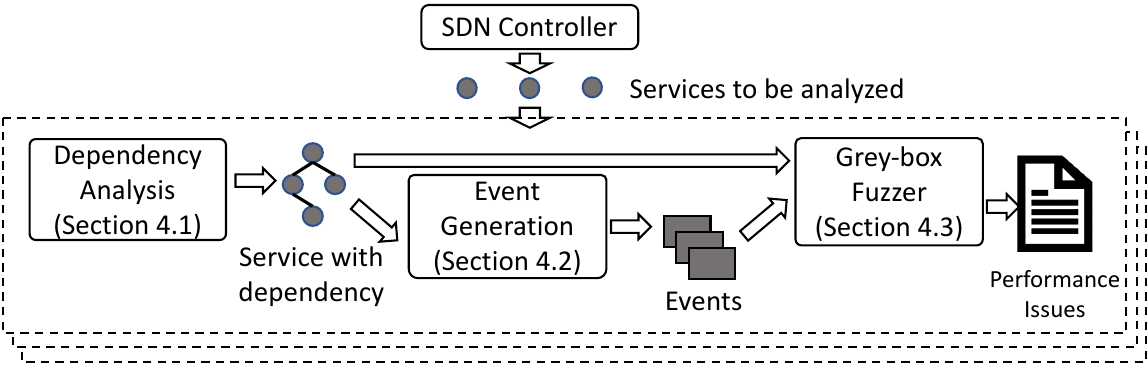}
    \tightcaption{ A high-level overview of \systemname. }
    \label{fig:decoupling}
    \vspace{-0.2cm}
\end{figure}

We adopt a fuzzing-based workflow to address our problem. 
Fuzz testing~\cite{fuzzing, aflfast, Manes19, Godefroid20} is a randomized input generation technique that effectively finds software bugs and security vulnerabilities in large and complex systems. 
However, we cannot apply existing fuzzing techniques directly.
We start by describing the design space for fuzzing and argue why strawman solutions do not work. Then, we describe our design choices to make this problem tractable.
We then present our end-to-end workflow, shown in Figure~\ref{fig:decoupling}.

\mypara{Design space and challenges} At a high level,
any fuzzing workflow entails the following choices that impose different trade-offs between fidelity, scalability, and manual effort:

\begin{itemize}
    \item {\em Granularity of code access:} One extreme is ``black-box'' fuzzing~\cite{fuzzing} with access only to the input/output of the system under test. At the other extreme, we have ``white-box'' fuzzing~\cite{sage}, which inspects source code to analyze state and execution paths.  Black-box approaches scale well but are imprecise, while white-box approaches are precise but do not scale to a complex codebase. A middle ground is ``grey-box''~\cite{aflfast}  fuzzing (e.g.,  AFL~\cite{afl} and libFuzzer~\cite{libfuzzer}), which uses lightweight instrumentation to get feedback from the test execution to guide input generation.
    \item {\em Granularity of  inputs:}
     Fuzzers can generate inputs in different representations, which entails a trade-off between the quality and the amount of domain knowledge that must be captured. In the simplest case, we send a raw bitstream. At the other extreme,  we can directly generate internal data structures for classes. There are also intermediate options; e.g., sending semantic-aware OpenFlow messages.


    \item {\em Granularity of system-under-test:} At one end, we can consider a monolithic view of the entire system, but this is also the least scalable. Alternatively, we can analyze individual classes, but we may miss out on vulnerabilities triggered by inter-class dependencies.

\end{itemize}

 A strawman workflow is to use \blackbox SDN fuzzers like Delta~\cite{delta} to
  generate OpenFlow message inputs to ONOS and check if some
  message(s) cause performance issues.
However, given the large input space, this approach does not work well, and most inputs are not relevant for \stateful scenarios. Consider Figure~\ref{fig:castor};
the function \texttt{add}
is called if and only if an OpenFlow message is received
by ONOS and the packet contains an ARP payload with the operation code \texttt{0x1} or \texttt{0x2} (Lines~\ref{fig:castor:payload}--\ref{fig:castor:check}).
Indeed,  we tried using Delta to randomly sample ten thousand OpenFlow messages. Of these, Delta produced 1140 OpenFlow
messages with ARP payloads. Only 13/1140 packets trigger the \texttt{add}
method and increase the size of \texttt{addressMap}. To increase the
execution cost of the \texttt{ARPService}, the fuzzer needs to generate more than
900 OpenFlow messages with valid ARP payloads.

\myparagraph{Design choice 1: Performance-oriented grey-box fuzzing.}  \SPIsShort require us to generate a sequence of relevant messages. The search space of individual messages alone is large, and considering a sequence further increases the search space. Thus, black-box fuzzers are not directly applicable.
  Grey-box performance fuzzers, such as SlowFuzz~\cite{slowfuzz} or PerfFuzz~\cite{perffuzz}, are a more promising starting point to
 tame large search spaces by evolving inputs via feedback from program executions. However, the complexity and semantics of ONOS pose key challenges that we need to tackle.

\myparagraph{Design Choice 2:  Event sequences as inputs.}
 Having chosen a grey-box workflow, we next consider the input granularity. A naive solution is to use a raw bitstream, but this lacks protocol semantics, causing most inputs to be dismissed as garbage. Alternatively, using OpenFlow messages and relying on the controller to convert them into internal states for each service also proves impractical, as the space of possible messages is too large. To address these issues, we leverage a domain-specific insight. As mentioned in \S\ref{sec:background}, ONOS uses an event-based architecture where incoming OpenFlow messages trigger new events. \SPIsShort occur when a service $S$ reaches an internal state that makes handling an \emph{event} costly, with the state depending on all previously handled events. This allows us to make the problem more tractable by searching for a \emph{sequence of events} instead of OpenFlow messages; i.e., we search for a sequence of events $\feventlist = \fevent_1, \fevent_2, \dots, \fevent_N$ such that the processing time of event $\fevent_N$ exceeds a predefined threshold.


 \myparagraph{Design Choice 3: Dependency-aware modular analysis.}
With an event-based fuzzing workflow, we see an opportunity to improve the scalability of our analysis without compromising its fidelity. As mentioned in \S\ref{sec:background}, the controller is composed of services that handle specific event types. If a sequence of events $\feventlist = \fevent_1, \fevent_2, \dots, \fevent_N$ triggers a performance issue in a service $\fservice$ that handles the event $\fevent_N$, we only need to search over prefixes $\fevent_1, \fevent_2, \dots, \fevent_{N-1}$ that directly or indirectly affect $\fservice$. Thus, we can reduce the search space by focusing on each service $\fservice$ individually, generating events handled by $\fservice$ or by any service interacting with $\fservice$ that impacts its state. This modular analysis is feasible due to our choice of a grey-box, event-driven workflow; a black-box approach or using packet/OpenFlow messages as inputs would require a monolithic analysis of the controller. We define a service as \emph{analyzable} if it registers at least one type of event. Therefore, we reformulate our problem to take as input a list of services to be analyzed, selected from the set of analyzable services, instead of analyzing the entire controller code at once.

\mypara{Overview} Combining these design choices above, we have the following end-to-end workflow, as depicted in Figure~\ref{fig:decoupling}. For each service to be analyzed, we first compute its dependency set using static analysis. Then, for each service and its dependency, {\systemname} uses  event  generators and performance fuzzing to generate event sequences of interest that can trigger potential \SPIsShort. Finally, we validate these vulnerabilities by reconstructing OpenFlow message sequences that will trigger the fuzzer-identified event sequences. Note that these design choices naturally dovetail into each other to enable our analysis to be tractable; e.g., the modular decomposition would not be possible without a grey-box event-based workflow.  To realize this solution in practice, we still need to address a number of system design and implementation challenges that we address in the following sections.

%% file: files/detailed-design.tex
\section{Detailed Design}\label{sec:design}
Next,  we describe the detailed design of {\systemname}. 



\begin{figure}[!tb]
    \centering
    \includegraphics[width=0.41\textwidth]{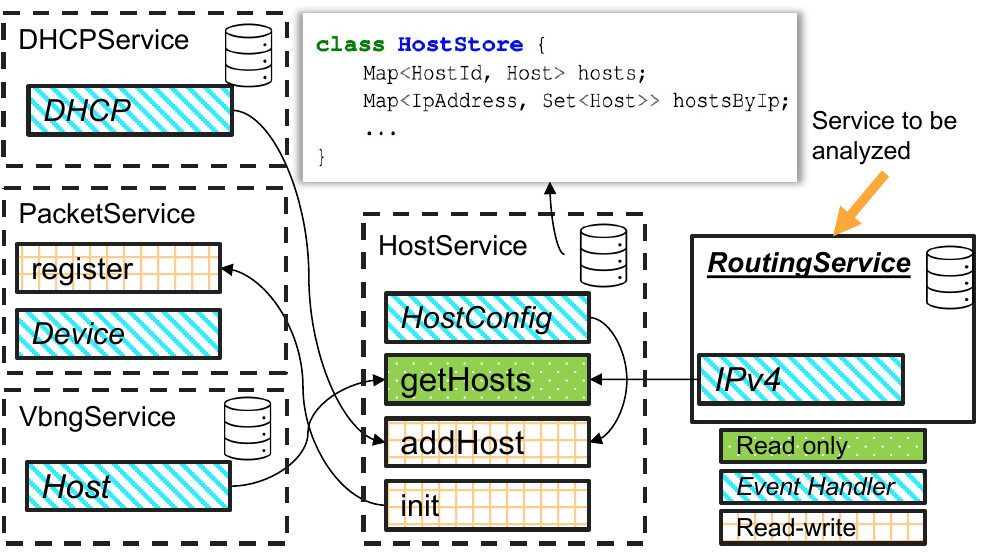}
    \tightcaption{Interactions between different services through function calls in ONOS. }
    \label{fig:call-graph}
    \vspace{-0.2cm}
\end{figure}


\subsection{Identifying Service Dependencies}\label{sec:design:decomposition}



A core benefit of \systemname{}'s design decision to search over event sequences is that it enables modular analysis instead of a monolithic analysis.
Specifically,  we can separately analyze each service in ONOS
to uncover \SPIsShort in that service.

Analyzing a service $\fservice$ involves searching for sequences
$\fevent_1, \fevent_2, \dots, \fevent_N$ of some fixed length $N$ such that $\fservice$ is an event handler
of $\fevent_N$.
Since we are interested in event sequences that trigger a performance issue when $\fservice$ is handling $\fevent_N$, we only care about events $\fevent_1, \fevent_2, \dots, \fevent_{N-1}$ that can affect the performance of the handler of
$\fevent_N$. Note that the event sequence includes events  handled by some other services
$\fservice'$ such that $\fservice'$ affects the internal state of the service $\fservice$.
We call the set of such services $\fservice'$ as the \emph{\statedependency set} of $\fservice$.
But how do we determine the \statedependency set?

Observe that the state of $\fservice$ may be manipulated by another service $\fservice'$ that calls a function in $\fservice$. Additionally, $\fservice$ may call a function in $\fservice''$, query the state of $\fservice''$, and then update its own internal state. Therefore, we would put $\fservice'$ and $\fservice''$ in the dependency set of $\fservice$, and then we also have to consider services that affect the states of $\fservice'$ and $\fservice''$ {\em  transitively}. 

One way to compute the \statedependency set is to include all services that can reach the analyzed service through function calls or be reached by it.
 Figure~\ref{fig:call-graph} presents a simplified
call graph for a subset of services.
Each edge represents a function call pointing from the callee
to the caller. In this example, the \statedependency set of \texttt{RoutingService}
based on this call graph would include \texttt{VbngService}, \texttt{HostService}, \texttt{PacketSerivce}, and \texttt{DHCPService}.


However, the call graph approach may include services that do not affect the state of the analyzed service. For instance, \texttt{VbngService} does not modify the state of \texttt{HostService}, since it only calls a read-only function \texttt{getHosts}; therefore, it cannot indirectly affect the state of \texttt{RoutingService}. We want the dependency set to be as small as possible to reduce the search space for analyzing a given service.

To this end, we use a refinement that reduces the search space without sacrificing analysis fidelity.
First, for each event handler, we compute a set of \emph{read} and a set of \emph{write} objects accessed
by the handler. We use this set to exclude services that do not affect the same state object
of the analyzed service \emph{while processing events}.
For example, the state of \texttt{HostService}
is not affected by \texttt{VbngService} and \texttt{RoutingService} because
\texttt{getHosts} only reads from the \texttt{HostStore}.
Additionally,
generating events for \texttt{PacketService} will not
affect the state of \texttt{HostService} because the \texttt{Device} event handler does not access the \texttt{HostStore} state object
at all.

Formally, our algorithm for computing the dependency set $\fdependency$ of a service $\fservice$ is as follows:
\begin{enumerate}
\item Initialize a set $R$ of state objects \emph{read}
by the event handlers of the analyzed service $\fservice$ and initialize
 $\fdependency$ to $\{ \fservice \}$.
\item\label{enum:dep} For each service $\fservice'$ that can reach the
analyzed service $\fservice$ through function calls or be reached by it:
\begin{enumerate}[label=\alph*.]
    \item Compute two sets, $R_{\fservice'}$ and $W_{\fservice'}$, containing state objects \emph{read} and \emph{written} by its event handlers, respectively.
    \item If $W_{\fservice'}\cap R$ is not empty, update $R\leftarrow R\cup R_{\fservice'}$ and $\fdependency \leftarrow \fdependency \cup \{ \fservice' \}$.
\end{enumerate}
\item If the dependency set $\fdependency$ is updated, go back to Step~\ref{enum:dep}.

\end{enumerate}

With this optimization, the \statedependency set of \texttt{RoutingService} now 
only includes \texttt{HostService} and \texttt{DHCPService}. The set
excludes \texttt{VbngService} and \texttt{PacketService} because
the event handlers from them do not write to any state objects read
by the \texttt{RoutingService}.

\begin{figure}[!tb]
    \centering
    \includegraphics[width=0.41\textwidth]{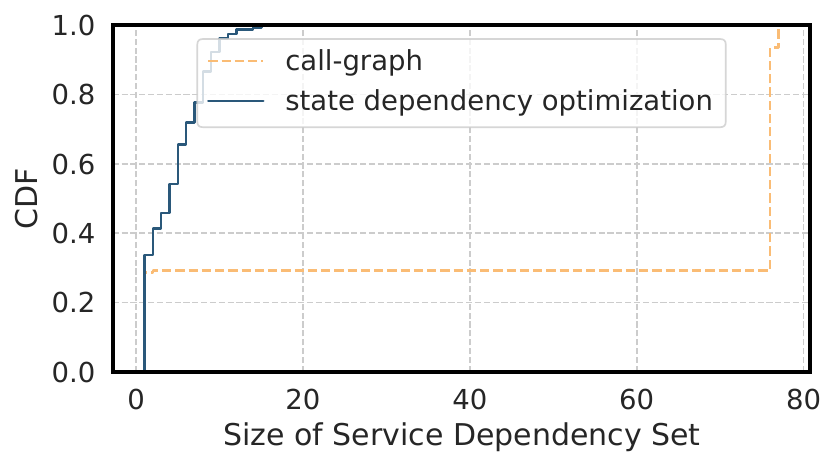}
    \tightcaption{CDF of \statedependency set sizes computed by the two algorithms across the 157 services analyzed in ONOS. A smaller size is better: the state-dependency optimization reduces the size of the dependency sets.}
     \vspace{-0.2cm}
    \label{fig:decomposition}
\end{figure}

As evidence of this optimization,
Figure~\ref{fig:decomposition} plots a CDF of the  service dependency sets across the
157 services in  ONOS. A naive call graph-based approach would have included over 75
dependent services for $\approx$ 70\% of the services. In contrast, our  state-dependency optimization results in a median of
4 and a maximum of 15 services in the dependency set. 

\subsection{Event Generation}\label{sec:design:event}

\begin{figure}[!tb]
\begin{minted}{Java}
class HostEvent {
    enum Type {
        HOST_ADDED, HOST_REMOVED
    };
    Host host;
    Type type;
}
class Host {
    String name;
    public Driver(String name{
      this.name = checkNotNull(name);
} }
\end{minted}
  \tightcaption{Simplified version of \texttt{HostEvent}.}
  \label{fig:host-event}
  \vspace{-0.3cm}
\end{figure}

Recall that analyzing a service $\fservice$ for \stateful \performance issues requires searching over \emph{event sequences} corresponding to events handled by any service in the \statedependency of $\fservice$. We decide to use generator functions for randomly sampling event objects. A generator for an event of type $T$ is a function $Random \rightarrow T$, where $Random$ is a source of randomness. This approach has been successfully applied by property testing tools such as Quickcheck~\cite{quickcheck, junit-quickcheck}.

In ONOS, events consist of data structure with multiple fields. For example, Figure~\ref{fig:host-event} shows a simplified version of \texttt{HostEvent}. The \texttt{HostEvent} contains two fields \texttt{host} with type \texttt{Host}, and \texttt{type} with type \texttt{Type}. \texttt{Host} is a data structure with one field \texttt{name} (type \texttt{String}). To randomly sample a \texttt{HostEvent}, we must randomly generate its fields recursively. So, we also need a generator for the type (\texttt{Type}), \texttt{Host}, and the name (\texttt{String}). To generate all event types, we need to be able to generate all fields recursively.

\begin{figure}[!tb]
\begin{minted}{Java}
class Generator {
  Object generate(Class type, Random rnd) {
  if (type == Integer.class) {
    return rnd.nextInt(); // random value
  } else if (...) { 
    ... /* other primitive types */
  } else { // object type
    Constructor c = type.getConstructor();
    Object o = c.newInstance();
    for (Field field: o.getFields()) {
      Object val = generate(field.getType(), rnd);
      field.set(val); // random value
    }
    return o;
}}}
\end{minted}
  \tightcaption{A simple type-based object generator that samples random \texttt{Object} instances given any \texttt{type}.}
  \label{fig:generator}
  \vspace{-0.2cm}
\end{figure}

By default, {\systemname} provides a type-based event generator that generates
events purely based on the type of each field~\cite{quickcheck, hotfuzz}.
Figure~\ref{fig:generator} presents the pseudocode of a type-based
object generator. The generator generates objects recursively
based on the type of each field. The automated approach is crucial to be able to quickly generate many types of events, but it has some limitations. In particular, events or other contained objects, when generated with unrestricted values for their fields, may violate certain constraints that the controller expects to be satisfied. Thus, the
type-based event generator may generate events that are \emph{invalid}.

Broadly, we identify two types of validity constraints:

\begin{figure}[!tb]
\begin{minted}{Java}
class HostEventGenerator {
  List<Host> generatedHosts;
  HostEvent generateHostEvent(Random rnd) {
    Type type = rnd.choose(Type.values());
    if (type == Type.HOST_ADDED) {
      Host host = generateHost(rnd);
      generatedHosts.add(host);
      return new HostEvent(host, type);
    } else if (type == Type.HOST_REMOVED) {
      Host host = rnd.choose(generatedHosts);
      generatedHosts.remove(host);
      return new HostEvent(host, type);
    }
  }
  Host generateHost(Random rnd) {
    ... // type-based random sampling
} }
\end{minted}
  \tightcaption{Simplified version of
  \texttt{HostEventGenerator}, which
   maintains inter-event constraints---hosts cannot be removed unless they have been previously added.
  }
  \vspace{-0.3cm}
  \label{fig:host-event-generator}
\end{figure}



\begin{itemize}
    \item \textbf{Intra-event constraints:} These
specify the internal constraints in an event. For example, in
Figure~\ref{fig:host-event} the \texttt{name} field of a \texttt{host} object should not be \texttt{null}; the constructor enforces this by calling a helper function \texttt{checkNotNull} which
will raise an exception if \texttt{name} is null.

\item \textbf{Inter-event constraints:} These are properties that must hold across multiple events.
There are two types of \texttt{HostEvent}, a \texttt{HOST\_ADDED} event is posted when a new
host is attached to the network, and a \texttt{HOST\_REMOVED} event is posted
when a connected host disconnects from the network. An inter-event
constraint is that a \texttt{HOST\_REMOVED} event is valid if and only
if the corresponding \texttt{host} has been added to the network and has not been removed.
\end{itemize}

In general, automatically generating such constraint-aware data structures is hard~\cite{langfuzz}.  While type-based event generators can be 
 used automatically for such events, they run the risk of generating invalid events. That is, either (a) the service handlers exit with an
error message without exercising meaningful behavior, or (b) the search for \SPIsShort may result in false positives.  


\begin{figure}[!tb]
\centering
    \includegraphics[width=0.42\textwidth]{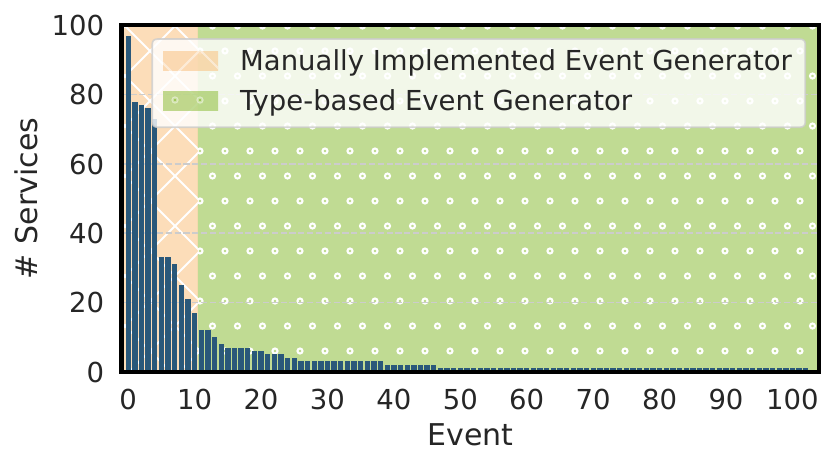}
    \vspace{-0.3cm}
    \tightcaption{For each event type in ONOS (X-axis), this plot shows the number of services whose states are affected by that event type (Y-axis). \systemname{} uses custom event generators for the top 10 critical events and automates the generation for the rest based on type information.}
    \vspace{-0.3cm}
    \label{fig:service-coverage}
\end{figure}


As a pragmatic compromise between manual effort and  coverage,  we choose  to manually implement 
 generators for only the most \emph{critical} of events,
and use automatic type-based generation for the rest.
We identify critical events by counting the number of distinct services whose states are affected by the event.
Figure~\ref{fig:service-coverage} shows that a small number of events affect the state of most services. Therefore, we write manual constraint-aware generators
(similar to  \texttt{HostEventGenerator} in Figure~\ref{fig:host-event}) for the top 10 events.



\subsection{Putting It Together}
\label{sec:design:fuzzing}

To generate event sequences for finding \SPIsShort, we start by identifying the service dependency set $\fdependency_\fservice$ for each service $\fservice$ (\S\ref{sec:design:decomposition}). Next, \systemname{} determines the event types $E_\fservice$ covering all event handlers registered by the services in $\fdependency_\fservice$. Using event generators (\S\ref{sec:design:event}), \systemname{} searches for event sequences $e_1, \ldots, e_N$, ensuring that each event’s type is in $E_\fservice$—i.e., each event is handled by at least one service in $\fdependency_\fservice$. The search runs for a time budget $B$ to find sequences where the performance cost of handling $e_N$ exceeds a threshold $t_{\max}$. Parameters $N$ and $B$ are determined based on available compute resources (\S\ref{sec:experiment}). The threshold choice is discussed later in our experiments.


\systemname{} performs the search by combining ideas from PerfFuzz~\cite{perffuzz}, a mutation-based grey-box performance fuzzer, and Zest~\cite{zest}, which applies mutation-based grey-box fuzzing to domain-specific input structures using generator functions.

\systemname{}'s algorithm for fuzzing a service $\fservice$
and its dependencies $\fdependency_\fservice$ with relevant event types
$E_\fservice$ combines performance and semantic fuzzing, as follows:

\begin{enumerate}
    \item Initialize a set $\Q$ with a randomly generated event sequence $\feventlist_0 = e_1, ..., e_N$, where the type of each event $e_i$ is chosen randomly from $E_\fservice$, and the event is randomly sampled via its corresponding event generator (\S\ref{sec:design:event}).
    \item Initialize a map $maxCounts$, which tracks the maximum execution cost observed at each program branch, by sending the event sequence $\feventlist_0$ to ONOS and monitoring the execution cost when processing $e_N$.
    \item \label{item:fuzz-mutate} Pick a random event sequence $\feventlist$ from $\Q$ and mutate it into a new event sequence $\feventlist'$ using the \emph{semantic fuzzing}~\cite{zest} approach, as described above.
    \item Send $\feventlist'$ to the services in $\fdependency_\fservice$
    and collect its execution instruction trace when processing the last event in $\feventlist'$.
    \begin{enumerate}[itemsep=0pt,label=\alph*.]
        \item If the total execution path length is greater than $t_{\max}$, then flag $\feventlist'$ as a potential issue.
        \item Otherwise, cumulate the element-wise execution cost of each program branch when processing the last event, and update the corresponding entry for each branch in $maxCounts$ if the value is greater.
        \item If any item in $maxCounts$ was updated, then add $\feventlist'$ to $\Q$. Otherwise, discard $\feventlist'$.
        \item If the time budget $B$ has expired, then stop fuzzing. Otherwise, go back to step \ref{item:fuzz-mutate}.
    \end{enumerate}
\end{enumerate}

Note that potential issues flagged by this process still require validation. First, an event $e_i$ in the flagged event sequence might be invalid if it violates intra-event or inter-event constraints (\S\ref{sec:design}), resulting in a \emph{false positive}. Second, finding a sequence of valid events that trigger high execution costs does not necessarily mean that the same effect can be caused by external OpenFlow messages. Currently, we manually translate an event sequence into an OpenFlow message sequence, and automating this step is a natural direction for future work. We label an issue as a \emph{true positive} if we can (manually) trigger the performance issue in a network emulation.

%% file: files/implementation.tex
\section{Implementation}\label{sec:implementation}

\myparagraph{Performance fuzzing and state reset.} 
 We implement the performance fuzzer in Java and Kotlin~\cite{kotlin} on top
of the JQF~\cite{jqf}  framework, which we extend to support
performance fuzzing~\cite{perffuzz}(\S\ref{sec:design:fuzzing}).
{\systemname} uses ASM~\cite{asm} and ByteBuddy~\cite{bytebuddy}
to instrument ONOS to collect the performance costs of events. 
Performance fuzzing requires that the state of the analyzed system 
can be reset easily across fuzzing executions.
A naive option for fuzzing is to launch a new instance for each
 execution. However, starting an instance of ONOS 
is slow (e.g., 30 seconds on a
laptop with 6-core and 32GB memory).  Alternatively,
 reusing an instance across fuzzing runs is not viable
  as the state is impacted by previous events. We
also considered resetting service state by instantiating a new
service object and discarding the old one. However, services
such as \texttt{StorageService} implement distributed persistent
storage. This is not only slow to launch but also persists the
state to a local file system and does not actually reset state.
  
Our approach to enabling state reset is to leverage {\em mock services}  provided by developers for unit tests. For example, the distributed store in the fuzzing harness can be replaced with a mock in-memory store, whose state can be reset using APIs like \texttt{clear} or \texttt{reset}. Although this prevents us from identifying performance vulnerabilities in the distributed store itself, it allows us to analyze many services that rely on the store instead.

\begin{figure}[tb]
    \includegraphics[width=0.42\textwidth]{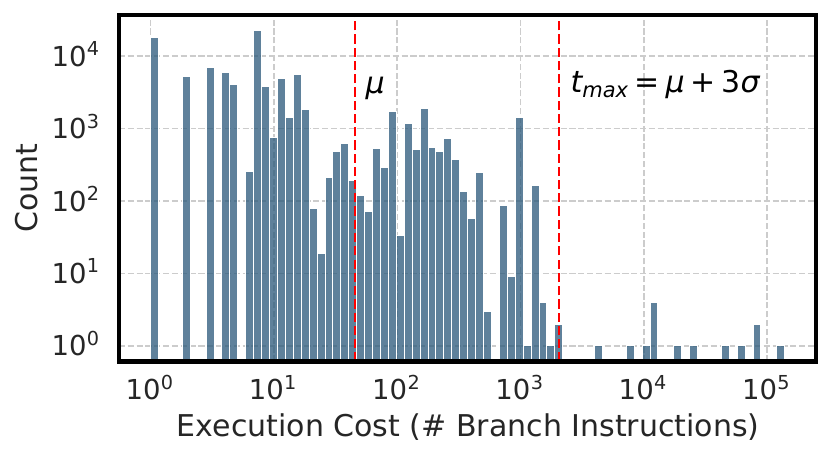}
    \tightcaption{Distribution of the per-event execution cost in a normal-workload emulation environment of ONOS ($\log$-$\log$ scale). We set the threshold at three standard deviations higher than the mean normal operation cost.}
    \label{fig:oracle}
     \vspace{-0.2cm}
\end{figure}

\myparagraph{Alerting threshold.}
Given a sequence of events generated by {\systemname}, we need to determine if the events trigger a potential \SPIShort. 
We use a data-driven threshold selection. First, we build a simple emulation environment using Mininet~\cite{mininet} with 4 hosts and 2 switches. We use \texttt{ping} utility to generate data plane packets and monitor execution costs of event handling in ONOS using JVM bytecode instrumentation.\footnote{One potential concern is that the processing time may depend on the specific deployment and topology size; i.e., is threshold based on a small topology relevant. We believe that this baseline is still useful as it indicates potential scalability bottlenecks inside the controller implementation.}  
As mentioned previously, we replace the distributed storage with in-memory storage to achieve an efficient state reset in ONOS. Note that the implementation of in-memory storage is much simpler than the distributed storage, and to avoid setting an unrealistic high threshold, we disable the instrumentation of the distributed storage in ONOS. We ran the emulation environment for 20 minutes, which resulted in 93,788 events being observed for ONOS. 
Figure~\ref{fig:oracle} shows the histogram of the per-event execution cost for  ONOS. We compare the execution cost of the last event generated by {\systemname} with the events generated in the simulation environment using z-score where $Z(x)=\frac{x-\mu}{\sigma}$ (i.e., the number of standard deviations above the mean). We use a threshold z-score of three
\footnote{The three-sigma rule based on the empirical rule in statistics states that for a normal distribution, approximately 99.7\% of the data lies within three standard deviations from the mean. Thus, any value with a z-score greater than 3 is considered abnormal.}: if $Z(x) > 3$, then we flag a potential a \SPIShort.\footnote{The outliers in Figure~\ref{fig:oracle} that have a cost higher than the threshold only appear during initialization; these are not considered as performance issues.}

\myparagraph{Validation and strategy reconstruction.}
For each potential vulnerability reported by {\systemname}, we want to verify if it actually represents a true vulnerability in ONOS.  Our insight here is that {events in 
the SDN controller provide useful information about events in the 
network}. For example, a \texttt{DeviceEvent}
represents that the status of a device is updated, which contains 
the type of update and detailed information about the device.
Similarly, a \texttt{Link} event represents the link update. Given 
a sequence of \texttt{DeviceEvents} and \texttt{LinkEvents}, we 
can dynamically reconstruct the topology. With this insight, 
 we can provide hints for a reconstruction strategy, including network topology information (i.e., hosts, switches, links) 
and network actions such as  OpenFlow messages (e.g.,  topology and 
configuration updates). We replay this sequence in network emulation using Mininet~\cite{mininet}. 


%% file: files/experiments.tex
\section{Evaluation} \label{sec:experiment}

\begin{table*}[!htb]
    \tightcaption{Summary of  performance issues identified by {\systemname} and baselines in  ONOS. Each row shows
    the affected class, a description of the issue, the source of OpenFlow messages that can trigger the issue, the smallest empirical sequence length  to uncover the issue}
    \label{tab:algo_result}
    \input{figs/algo_result.tex}
     \vspace{-0.4cm}
\end{table*}




We evaluate {\systemname} on ONOS v2.2.4~\cite{onos}. Our evaluation is focused on answering the following research questions.

\myparagraph{RQ1.} Is {\systemname} effective at identifying \SPIsShort in ONOS?

\myparagraph{RQ2.} How does \systemname{} compare to a traditional SDN fuzzer in identifying \SPIsShort?

\myparagraph{RQ3.} To what extent does the dependency-aware modular fuzzing technique help in identifying \SPIsShort{}?



For each service to be analyzed, we have two parameters to scope the analysis: (1) {\em time budget} ($B$) to run the analysis and (2) {\em sequence length} ($N$) of events to explore. With longer time and length, the fuzzer consumes more resources and has a greater chance of identifying \SPIsShort. However, longer sequence lengths also increase the search space. We configure {\systemname} to find a sequence of events with lengths  $N$=1, 100, 250, 500, 1000, and 2500. 

For each $N$, we allocate a budget $B$ of 1 hour to analyze each service. The fuzzer also uses results from previous sequence lengths as seeds, and the total fuzzing time of each service is 6 hours. We repeated each experiment 5 times, which led to a total of 4740 CPU hours (197 CPU days) per configuration. We conduct all of our experiments on Cloudlab VMs using 4 cores (2.4 GHz) and 4 GB memory for each service.
The fuzzer runs and reports the smallest N, where the z-score of the cost of handling the last event is greater than 3, or NULL if no such event was found 
as described in \Section\ref{sec:implementation}.

\subsection{RQ1: \SPIShort Detection with \systemname{}}


After fuzzing each of the 157 services with the above parameters, \systemname{} reported 11 potential issues, summarized in Table~\ref{tab:algo_result}. We manually analyze these 11 reports and find that 10 are true positives (which we name V1--V10) while one is a false positive (named F1). Out of the 10 true positives,  9 of these are truly {\em stateful} performance issues; i.e., they require a non-empty of sequence of events to set up a vulnerable state before the issue can be triggered. Only V8 can be triggered with a single event. We manually inspect F1 and identify that it relies on an automatically synthesized type-based event generator for \texttt{ControlMessageEvent}, which does not take into account some constraints and produces an invalid event (e.g. the maximum allowable size of a control-message list is exceeded); therefore, the issue cannot be triggered using OpenFlow messages.
We also verify that all performance issues can be triggered regardless of the implementation of the storage layer in ONOS.

\myparagraph{Validation/Replay.}
For each reported issue, we use Mininet
to manually reconstruct the issue.
We successfully replicated 9 issues in the
 emulated network.\footnote{We are not able to replicate V4 due to the another bug triggered
by the emulator (Case Study 2).
}

\myparagraph{Responsible disclosure.}
We have notified the ONOS developers and presented them with concrete end-to-end traces to reproduce our reported issues. 


\myparagraph{Classification.} We manually  classify the
10 performance issues along two dimensions: {\em source of the triggering event} and {\em algorithmic complexity}.
 First, we classify issues based on the types of sources that can generate key events to
trigger these issues:
  {\em host}, {\em switch}, {\em controller}. For example, any host connected to the network can generate \texttt{PacketIn} events with IPv4 payloads, so its source is classified to {\em host}. 
  A \texttt{PacketIn} event with LLDP payload can only be sent by switches, so its source is classified as {\em switch}. Some events can only be triggered by an SDN controller configuration update, and those events
  will have {\em controller} as the source.
  Second, we qualify the algorithmic complexity of
 the performance issue as a function of the number of events in the sequence. Specifically,
  we identify {\em high constant}, {\em linear}, and
  {\em exponential} patterns of {\em per-event} execution time for the trigger event. Note that per-event linear complexity translates to a cumulative performance cost of $\mathcal{O}(n^2)$ for $n$ events.


Table~\ref{tab:algo_result} presents a comprehensive listing of all issues discovered by {\systemname} and baselines.
  Out of 10 true positives,   2 issues
can be triggered from a malicious host, which is the most serious case; 7 issues
can be triggered from compromised switches;
1 issue can only be triggered by ONOS itself. While the latter is not a serious security risk, it may occur due to accidental misconfigurations.

The non-stateful issue V8 causes ONOS to perform a high- constant-cost execution; 8 issues cause ONOS to perform a computation whose per-event cost increases linearly with respect to the number of events generated; 1 issue (V4) causes ONOS to perform a computation whose cost increases exponentially with respect to the events or generated.





\begin{figure}[!tb]
\centering
    \includegraphics[width=0.42\textwidth]{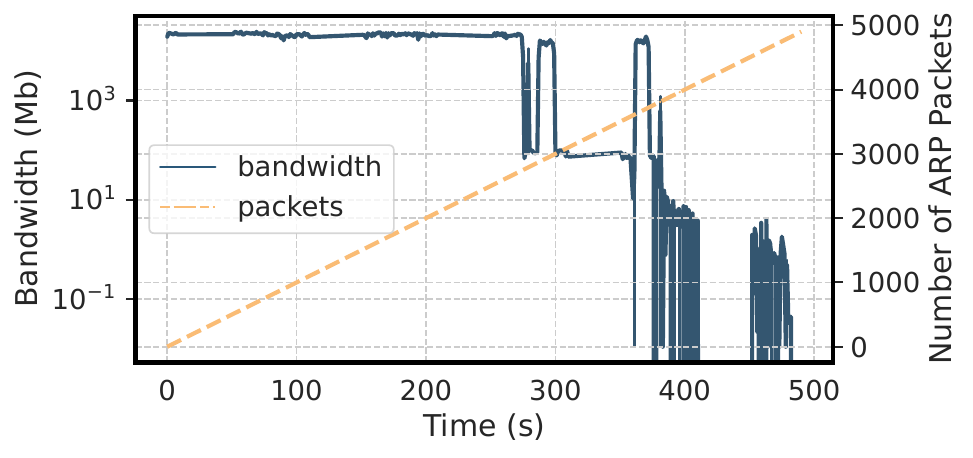}
    \vspace{-0.3cm}
    \tightcaption{The throughput seen by benign hosts drops significantly 300 seconds after the 
    ARP spoofing attack starts.}
    \vspace{-0.3cm}
    \label{fig:arp-bandwidth}
\end{figure}

\myparagraph{Case Study 1: Host-initiated stateful performance issue via spoofed ARP packets (V1).} This issue, in class \texttt{CastorArpManager}, can be exploited by any malicious hosts in the network. The root cause of the issue is depicted (highly simplified) in Figure~\ref{fig:castor}. The execution cost of the ARP-related service increases as more ARP records are added to an internal data structure. We manually reconstructed the OpenFlow messages that triggered this issue. As a proof-of-concept, we use Mininet~\cite{mininet} to create a simple network with three switches. Each switch connects to one host. We use one host to generate spoofed ARP packets and monitor the connectivity between the other two hosts. The malicious host generates 10 spoofed ARP packets per second to avoid the flooding attack.


We use iperf to measure the bandwidth between two benign hosts in the network, with results shown in Figure~\ref{fig:arp-bandwidth}. The bandwidth started at 27 Gbits/s, but dropped significantly after 270 seconds. To disrupt the network, the attacker only needed to create 3,000 fake ARP packets at a low frequency. This wasn't due to a data plane attack, as confirmed by a separate test without \texttt{CastorArpManager}. The \SPIShort increased ONOS's processing time for OpenFlow messages, affecting its throughput. OpenFlow messages containing LLDP data checked link liveness, but ONOS couldn't process them quickly enough during the attack, marking links as unavailable and impacting network bandwidth.


Moreover, it is hard to fix the issue with easy configuration patches or reboots. ONOS saves all ARP records in persistent storage, and it does not provide an interface to remove a single field unless the user removes the entire data store. In that case, other configurations will also be removed.

\begin{figure}[tb]
\centering
  \includegraphics[width=0.42\textwidth]{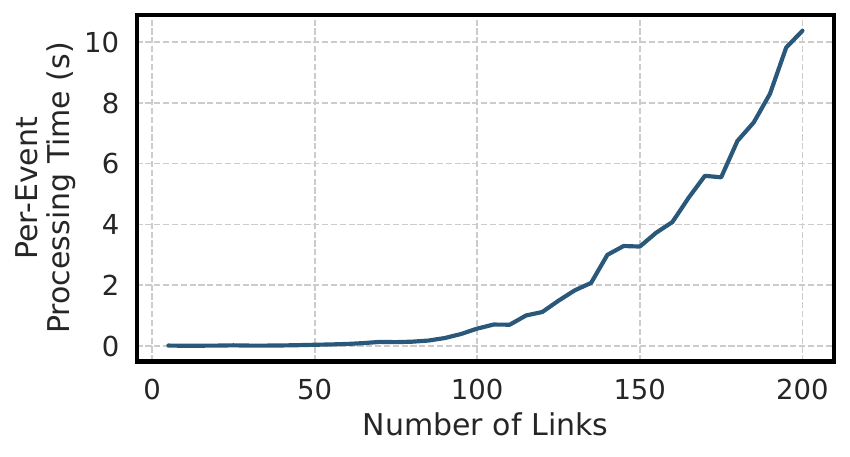}
  \tightcaption{Execution time of \texttt{AbstractGraphPathSearch}
  service increases exponentially with respect to the number of paths created in the
  network.}
  \vspace{-0.3cm}
  \label{fig:p4-attack}
\end{figure}

\myparagraph{Case Study 2: Exponential-time stateful performance issue induced by redundant links (V4).}
{\systemname} reports that
the execution cost of the \texttt{AbstractGraphPathSearch} method increases exponentially
with respect to the number of links in the network, in particular when there are \emph{redundant links}
between devices. This is incredibly subtle because the link graph is actually a multi-graph, and the path search algorithm degrades in the presence of multiple edges between the same pair of nodes. {\systemname} identifies this issue by generating
a topology with multiple redundant links.

In order to replay this issue, we used Mininet to generate a simulation network
with redundant links. Unfortunately, the simulation environment triggers an unrelated bug
in ONOS which hangs up the controller completely and stops processing any
OpenFlow messages from the data plane.

However, we are still able to trigger the issue that \systemname{} discovered by implementing a standalone service that can send messages to \texttt{TopologyService}. We use this service to generate a topology containing 5 devices, and then slowly add redundant links by sending appropriate messages. Figure~\ref{fig:p4-attack}
shows the performance of the \texttt{TopologyService}, which uses
\texttt{AbstractGraphPathSearch} to compute paths between nodes, with
respect to the total number of links created in the network.
This subtle case of redundant links in a multi-graph topology demonstrates that {\systemname} can identify hard-to-detect \stateful performance issues.

\subsection{RQ2: Compare to SDN Fuzzer}

In order to answer RQ2, we use a packet fuzzer adopted from Delta~\cite{delta} to fuzz ONOS for 12 hours. Delta is a state-of-the-art black box SDN fuzzer, which generates stateful OpenFlow messages. Since this fuzzer does not use instrumentation, we further instrument ONOS and measure the execution cost of each event triggered by the SDN fuzzer. 

Delta triggers over 2 million internal events in total and only 582 events whose z-scores are greater than 3. Except for the events generated during the bootstrap stage of ONOS similar to the simulation environment shown in Figure~\ref{fig:oracle}, Delta triggers 3 \SPIsShort (V2, V3 and V5) but does not uncover any new \SPIsShort not already identified by \systemname{}. Furthermore, Delta fails to detect subtle issues that alter the topology. Note that Delta generates OpenFlow messages continuously without resetting; therefore, any state changes are unintentional. It is thus non-trivial to isolate small message sequences that trigger an \SPIShort. From this experiment, we can conclude that \systemname{} is more effective than Delta in identifying \SPIsShort.


\subsection{RQ3: Sensitivity Analysis}

\begin{figure}[tb]
\centering
  \includegraphics[width=0.42\textwidth]{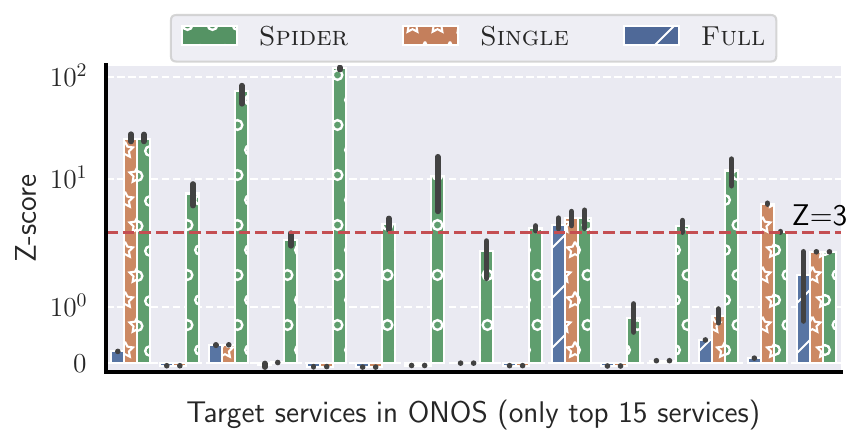}
  \tightcaption{The z-score of the target services reported by different fuzzers.}
  \vspace{-0.3cm}
  \label{fig:z-score}
\end{figure}

To evaluate the efficacy of {\systemname}'s dependency-aware modular fuzzing design, we compare two baselines {\sysempty} and {\sysfull}. {\sysempty} do not use dependency analysis and fuzz each service without any dependency, and {\sysfull} fuzzes each service with all services as dependencies. We use {\sysempty} and {\sysfull} to analyze all 157 services with the same configuration as {\systemname} and repeat the experiment 5 times.

\myparagraph{Bug reports.} As shown in Table~\ref{tab:algo_result}, all {\SPIsShort} reported by {\sysfull} and {\sysempty} are covered by {\systemname}, which consistently identifies all {\SPIsShort} across each repetition. {\sysfull} only reports one performance issue (V7), which can be identified easily by generating only one event with an IPv4 payload. {\sysfull} fails to identify all stateful performance issues that require more than one event to trigger the issue. We found that including all services as the dependency of the analyzed service greatly decreased the performance of the fuzzer because the fuzzer needs not only to take more time to initialize each service but also to explore a larger state space that is irrelevant to the analyzed service. {\sysempty} reports two performance issues (V1 and V8) and one false positive (F1). All three performance issues can be identified by only exploring the state space of the analyzed service. {\sysempty} cannot identify other performance issues because the search space has been artificially limited.

\myparagraph{Finding the worst case input.} Figure~\ref{fig:z-score} shows the z-score of the worst-case input of different services identified by different fuzzers across multiple runs. We only show services whose z-score exceeds 1 for at least one fuzzer. Not that there are 12 services whose Z-scores are greater than 3 because different services trigger the same \SPIShort. {\systemname} outperforms {\sysfull} in 14 out of 15 services. This confirms that it is important to use modular fuzzing to reduce the search space for the performance fuzzer. {\systemname} out performs {\sysempty} in 11 out of 15 services. {\sysempty} reports a higher z-score if the worst-case input can be constructed without exploring the state space of other services. However, it failed to construct complex state full input for other services. 
Our result shows that the dependency-aware modular design is critical in identifying \SPIsShort.

%% file: figs/algo_result.tex
\renewcommand{\arraystretch}{1.2}
\begin{footnotesize}
\begin{center}
    \begin{tabular}{ @{}c  p{16mm}  p{74mm}  c  c  c  c  c  c@{}}
    \toprule
     ID & Service & Description & Source& Smallest & {\systemname} & {\sysfull} & {\sysempty} & SDN-\\
     & Name & & & $N$ &  &  &  & Fuzz \\
     \midrule
     V1 & Castor
     &The execution cost of \texttt{CastorArpService} increases linearly
     with respect to the number of \texttt{OFPacketIn} with ARP payload received by the service.
     & host 
     & 2500
     & \checkmark
     & \xmark
     & \checkmark
     & \xmark
     \\
     \hline
     V2 & Neighbor Resolution
     & The execution cost of \texttt{NeighborResolutionManager} increases linearly
     with respect to the number of connect points in the network.
     & switch 
     & 50
     & \checkmark
     & \xmark
     & \xmark
     & \checkmark
     \\
     \hline
     V3 & Port Statistics
     & The execution cost of \texttt{PortStatisticsService} increases linearly
     with respect to the number of \texttt{OFPortStatisticsReply} messages
     received by the service.
     & switch 
     & 1000
     & \checkmark
     & \xmark
     & \xmark
     & \checkmark
     \\
     \hline
     V4 & Graph Path Search
     & The execution cost of \texttt{AbstractGraphPathSearch} service increases
     exponentially with respect to the number of links in the network.
     & switch 
     & 50
     & \checkmark
     & \xmark
     & \xmark
     & \xmark
     \\
     \hline
     V5 & My Tunnel App &
     The execution cost of \texttt{MyTunnelApp} increases linearly
     with respect to the number of hosts in the topology.
     & switch 
     & 50
     & \checkmark
     & \xmark
     & \xmark
     & \checkmark
      \\
     \hline
     V6 & VPLS &
        The execution cost of \texttt{VplsManager} increases linearly
        with respect to the number of interfaces configured in the SDN controller.
     & controller 
        & 50
     & \checkmark
     & \xmark
     & \xmark
     & \xmark
     \\
     \hline
     V7 & Links Provider & The execution cost
     of \texttt{NetworkConfigLinksProvider} increases linearly with respect to
     the number of port created for each switch.
     & switch 
     & 50
     & \checkmark
     & \xmark
     & \xmark
     & \xmark
     \\
     \hline
     V8 & Rabbit MQ & The \texttt{MQEventHandler} performs a costly
     computation while processing IPv4 packets.
     & host 
     & 1
     & \checkmark
     & \checkmark
     & \checkmark
     & \xmark
     \\
     \hline
     V9 & Router Advertisement & The execution cost of
     \texttt{RouterAdvertisementManager} increases linearly with respect to the number of
     interfaces created in the network.
     & switch 
     & 50
     & \checkmark
     & \xmark
     & \xmark
     & \xmark
     \\
     \hline
     V10 & Link Discovery & The execution cost \texttt{LinkDiscoveryProvider}
     increases linearly with respect to the number of switches in the network.
     & switch 
     & 1000
     & \checkmark
     & \xmark
     & \xmark
     & \xmark
     \\
     \hline
     F1 & Control Plane Manager & An invalid
     \texttt{ControlMessageEvent} causes high execution
     of the \texttt{ControlPlaneManager}. & N/A & N/A 
     & \checkmark
     & \xmark
     & \checkmark
     & \xmark\\
    \bottomrule
    \end{tabular}
\end{center}

\end{footnotesize}

%% file: files/related-work.tex
\section{Related Work}
\label{sec:related}

\myparagraph{SDN fuzzers.}
Existing black-box SDN fuzzers (e.g., Beads~\cite{beads} and Delta~\cite{delta})
generate packets based on an existing topology and
focus on logic protocol bugs in SDN controllers~\cite{beads, delta}.
Most
OpenFlow messages generated by the SDN fuzzer only explore
\emph{a small portion of the input space} of the SDN controller, and
many performance-sensitive services
are left untested using black-box SDN fuzzers.

\myparagraph{Other analysis of SDN controllers.}
Nice~\cite{nice} uses symbolic execution and model checking to
identify property violations.
ConGuard~\cite{conguard} and SDNRacer~\cite{sdnracer} use static analysis
to identify race conditions in the SDN controller. EventScope~\cite{eventscope}
focuses on missing event handlers in SDN applications, and AudiSDN~\cite{audisdn}
identifies inconsistent policies among different modules.
None of these efforts tackle \SPIsShort.

\myparagraph{Static code analysis for performance.}
Static performance analysis techniques (e.g., FindBugs~\cite{findbugs}, Clarity~\cite{clarity},
Torpedo~\cite{torpedo}) identify performance issues based on code patterns.
Unfortunately, specifying such patterns usually requires domain-specific knowledge and
many patterns of \SPIsShort are not described in existing tools. 

\myparagraph{Symbolic execution for performance analysis.}
Symbolic execution  (e.g., Castan~\cite{castan} and Wise~\cite{wise})
can be used to identify states with performance issues. However,
the state space of the program increases exponentially with respect to the size of the program.
Therefore, such techniques are still limited to analyzing small programs
and cannot handle the \emph{large state space} of the SDN controllers~\cite{symbolicsurvey}.

\myparagraph{Languages for performance analysis.}
Performance modeling languages such as RAML~\cite{raml} provide an estimation of the program complexity.
However, translating the existing SDN controller implementation into such languages is a challenge.
Similar to static performance analysis, performance modeling languages cannot model the existing
\emph{complex code base} of the SDN controller, such as reflection and runtime code generation.

\myparagraph{Trace-driven analysis.}
Dynamic performance monitors (e.g., Freud~\cite{Freud} and PerfPlotter~\cite{perfplotter})
collect execution traces and produce an algorithmic complexity
estimate~\cite{Freud, perfplotter}.
However, if the traces used for modeling (typically of common-case workloads) do not cover the (likely rare) \SPIShort patterns, such tools will not be able to uncover \SPIsShort.


\myparagraph{Fuzzing stateful network protocols.} Network fuzzers  use protocol specifications~\cite{snooze, kif, restler} or try to infer protocols automatically~\cite{pulsar, polyglot, aflnet}. These focus on protocol bugs or correctness issues, rather than \SPIsShort.  

%% file: files/discussion.tex
\section{Conclusion and Future Directions}

In some ways, our effort is a proof-by-construction of the viability of a seemingly intractable program analysis problem: uncovering deep semantic stateful performance issues in large and complex software.    
 We conclude by discussing extensions, limitations, and lessons.

 There are three immediate extensions. First,    capturing the semantic constraints in the top-10 events manually adds a lot of value. Thus, we can increase coverage by making the type-based generation more semantic-aware.  Second, we can make the reconstruction and validation process more automated (e.g., via program synthesis)  using \systemname's hints.
 Third, we need a way to also find issues in distributed components such as the state store for ONOS.


  Finally, our experience sheds light on the benefits of domain-specific insights in fuzzing and of design for testability.
  On a positive note,  the presence of mock services and unit tests simplified our implementation. At the same time, the lack of semantic-aware constructors made event generation hard. An interesting direction for future work is to discover such domain-specific invariants and provide hints to developers on how they can   support   fuzzing  workflows.

\section*{Acknowledgment}

We would like to thank the anonymous reviewers for their feedback. This research was supported in part by seed funding from CMU’s CyLab, the Cylab Future Enterprise Security Initiative, and the NSF grant CNS-2132639. 